%% file: lep_susy.tex
\documentclass[11pt]{article}
\usepackage{moriond,epsfig}

\input{kjtex.tex}

\def\Journal#1#2#3#4{{#1} {\bf #2}, #3 (#4)}

\def\PLB{{\em Phys. Lett.}  B}

\def\PREP{\em Phys. Rep.}

\def\be{\begin{equation}}
\def\ee{\end{equation}}
\def\bea{\begin{eqnarray}}
\def\eea{\end{eqnarray}}

\begin{document}
\vspace*{4cm}
\title{MSSM SUSY SEARCHES AT LEP2}

\author{ K. Jakobs}

\address{Institut f\"ur Physik, Universit\"at Mainz, Staudinger Weg 7, \\
55099 Mainz, Germany}

\maketitle\abstracts{
In the final run in the year 2000 the four experiments at the electron positron 
collider LEP have accumulated data corresponding to an integrated luminosity of 
more than 210 $\pbs$ per experiment at centre-of-mass energies up to 208 GeV. 
These data have been used to extend the search for supersymmetric particles in the 
new energy domain. In the present paper the results of searches for sfermions, 
charginos and neutralinos are described,
assuming the minimal supersymmetric standard model with R-parity conservation. 
In addition, the determination of a lower mass limit
of the lightest neutralino is discussed.}

\section{Introduction}
The search for particles predicted by supersymmetric theories~\cite{susy} (SUSY) is
one of the main motivations for the physics programme at LEP2. SUSY provides an 
elegant solution for several deficiencies of the Standard Model (SM). 
The minimal supersymmetric extension of the Standard Model (MSSM) 
predicts scalar partners for both helicity states of SM fermions (sfermions) and 
fermionic partners of SM gauge bosons (gauginos). Neutral (charged) gauginos will 
mix with the supersymmetric partners of the higgs bosons to form four neutral states
(neutralinos) and two charged states (charginos). Since
the predicted degeneracy between particles and their supersymmetric partners is not
observed supersymmetry must be broken.

In the searches described in the following a gravity mediated breaking of 
supersymmetry is assumed. In addition, the conservation of R-parity is assumed, 
which implies the pair production of supersymmetric particles, their decay into an
odd number of such particles and the stability of the lightest supersymmetric 
particle (LSP), which is assumed to be the lightest neutralino. 
The soft SUSY breaking introduces gaugino mass parameters $M_i$ into the theory, 
which are considered as free parameters together with the ratio of the vacuum 
expectation values of the two Higgs doublets, $\tanb$ and the mass parameter in the
Higgs sector $\mu$. For the interpretation of results in the MSSM the unification 
relation 
$M_1 = \frac{5}{3} \ \tan^2 \theta_W \ M_2$ is assumed. This relation is important for
fixing the masses and the field content of charginos and neutralinos. 

Results presented in this paper have been obtained using data taken during the run in 
the year 2000, combined with lower energy data, where appropriate. At center-of-mass
energies in the range between 203 and 208 GeV data corresponding to an integrated 
luminosity of more than 210 $\pbs$ have been collected by each experiment.
 
\section{Searches for Sfermions}
At LEP scalar partners of leptons and quarks are produced in pairs via the s-channel
annihilation diagram. For selectrons an additional contribution via  t-channel neutralino 
exchange appears.  
In the SUSY scenario 
considered,  sfermions decay predominantly into the corresponding SM fermion and the LSP, the
lightest neutralino $\chione$. These events are therefore characterised by two acoplanar 
leptons or jets accompanied by missing energy. For the first two generations the mass 
eigenstates correspond to the right- and left-handed states. Since production cross sections 
are lower for $\tilde{f}_R$ than for  $\tilde{f}_L$, the $\tilde{f}_R$ states are used to set 
conservative limits. In the third generation  $\tilde{f}_R$ and $\tilde{f}_L$ are not
necessarily mass eigenstates. The off-diagonal terms in the mass matrix are of the form 
$m_f ( A_f - \mu \cot \beta )$ for the stop, and $m_f ( A_f - \mu \tan \beta )$ for 
sbottom and stau and can cause substantial mixing. 

\subsection{Sleptons}
\begin{figure}
\begin{minipage}{10.5cm}
\epsfig{file=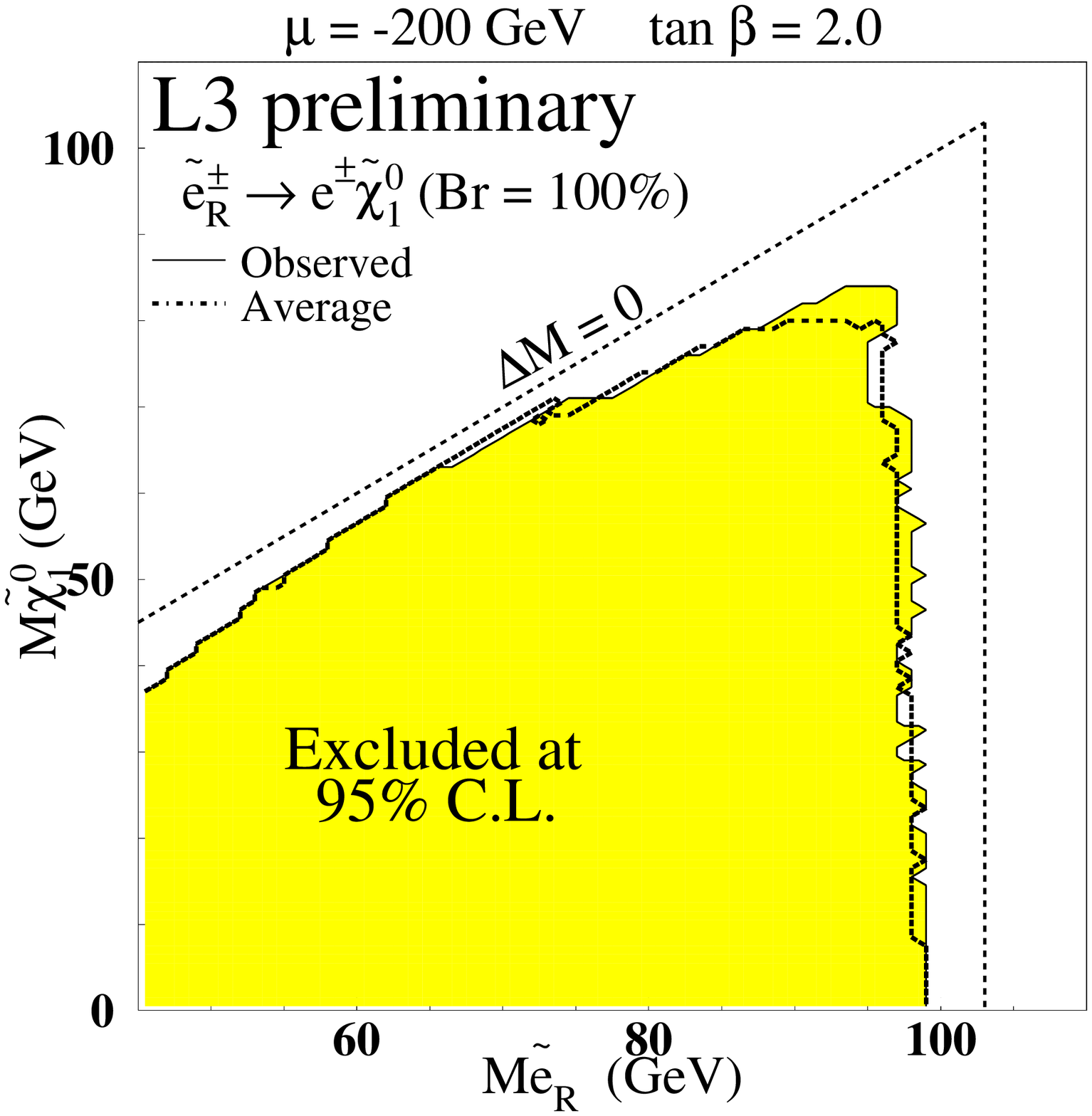,height=7.0cm}
\end{minipage}
\begin{minipage}{9.0cm}
\hspace*{-3.0cm}
\epsfig{file=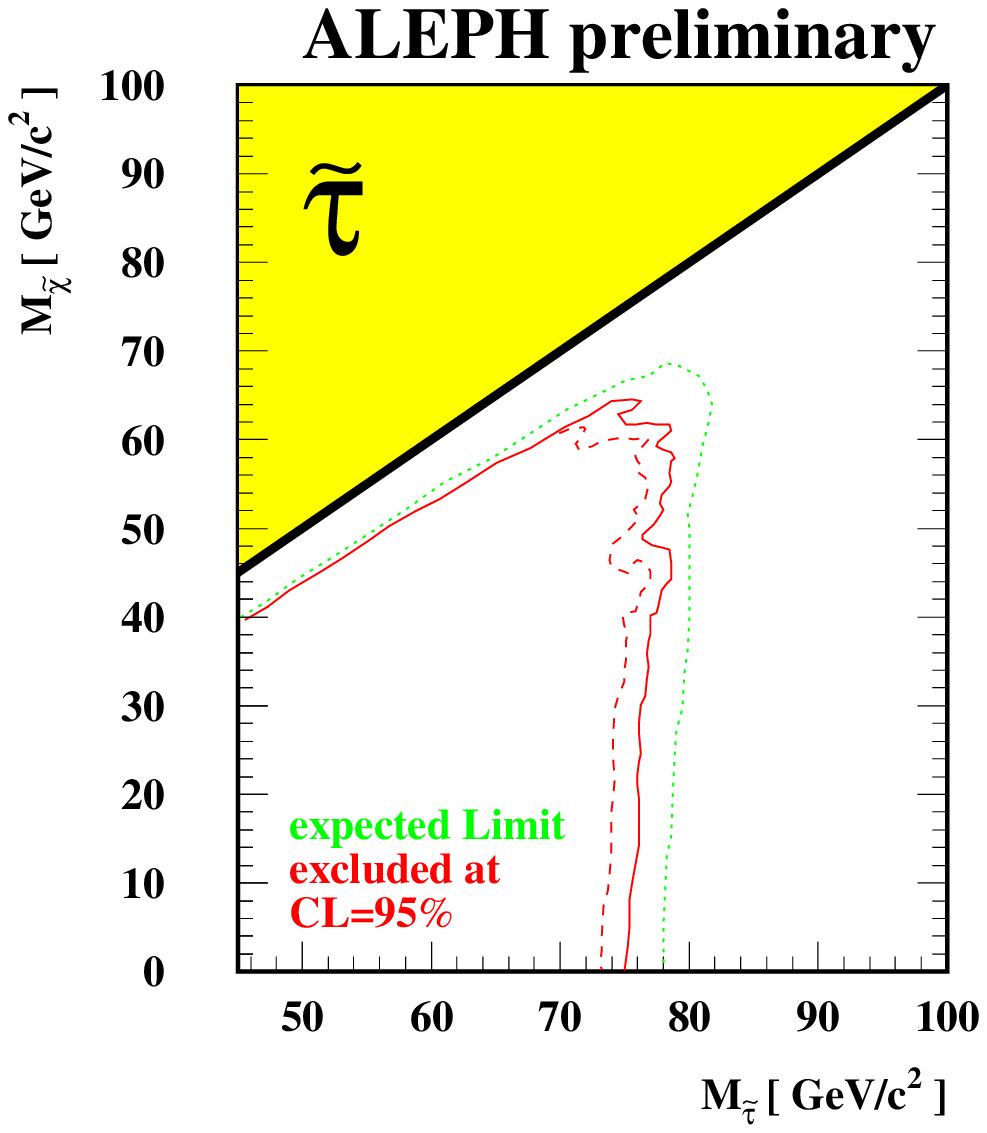,height=8.5cm}
\end{minipage}
\caption{Excluded regions (95\% C.L.) in the 
$M_{\slep} - M_{\chione}$-plane for selectrons (left) and staus (right). In the case of
staus the limit is shown for the no-mixing case (full line) and for the case of minimal coupling
of the lightest stau to the $Z^0$. 
In both cases 
also the expected limit is shown (full and dotted line for $\sel$ and $\stau$ 
respectively). 
\label{f:sleptons}}
\end{figure}

Searches for acoplanar leptons have been used to set exclude regions in the 
($M_{\slep} - M_{\chione}$)-plane. These masses are the main parameters in searches for the
first two generations: the slepton mass determines the production cross section, the mass 
difference $\delm$ to the LSP is related to the visible energy and thus to the selection
efficiency and background composition. Due to the t-channel neutralino exchange contribution, 
which can substantially enhance the production cross section, the results for selectrons
depend in addition to the mass parameters strongly on the choice of $\tanb$ and $\mu$. 
In Fig.\ref{f:sleptons} examples of the exluded regions in the mass plane are shown for 
selectrons (from the L3 collaboration)   
and staus (from the ALEPH collaboration). Both the regions excluded by the experimental 
data and the limits expected assuming the known efficiencies and background 
expectations are shown. 
For selectrons the exclusion is shown for $\tanb$=2.0 and $\mu$=-200 GeV. 

The production of staus depends in addition on the nature of the lighter stau, which 
is a composition of right- and left-handed states: 
$\stau_1 = \stau_L \cos \theta + \stau_R \sin \theta$. The most conservative limits are 
obtained chosing the mixing angle such that the production cross section is minimal, which
is close to the decoupling from the $Z^0$. In the Figure both the no-mixing case (full line)
as well as the case of minimal cross section (dashed line) are shown. 

The results of the four LEP experiments have been combined and preliminary limits using 
the full LEP data set have
been extracted. The results are summarized in Table~\ref{t:sfermions}. Assuming a mass value
of 40 GeV for the lightest neutralino, mass limits in the range between 87.1 (for staus) 
and 99.4 (for selectrons) have been extracted. There is good agreement between the 
observed and expected limits. This is also true in the case of the staus. The new data 
collected in the year 2000 by all LEP collaborations do not enhance the excess of stau
candidates which has been observed during the 1999 running. 
For example, the ALEPH collaboration 
observes 19 events in their high $\delm$ analysis for data collected in the 
year 2000, whereas 18.3 events are expected from know background sources. 

\subsection{Squarks}
Due to the large top mass, mixing in the stop sector can be large and the lightest 
stop can become the lightest squark. Also the mass of the sbottom could be low, in 
particular if $\tanb$ takes large values. 

\begin{figure}
\begin{minipage}{9.5cm}
\epsfig{file=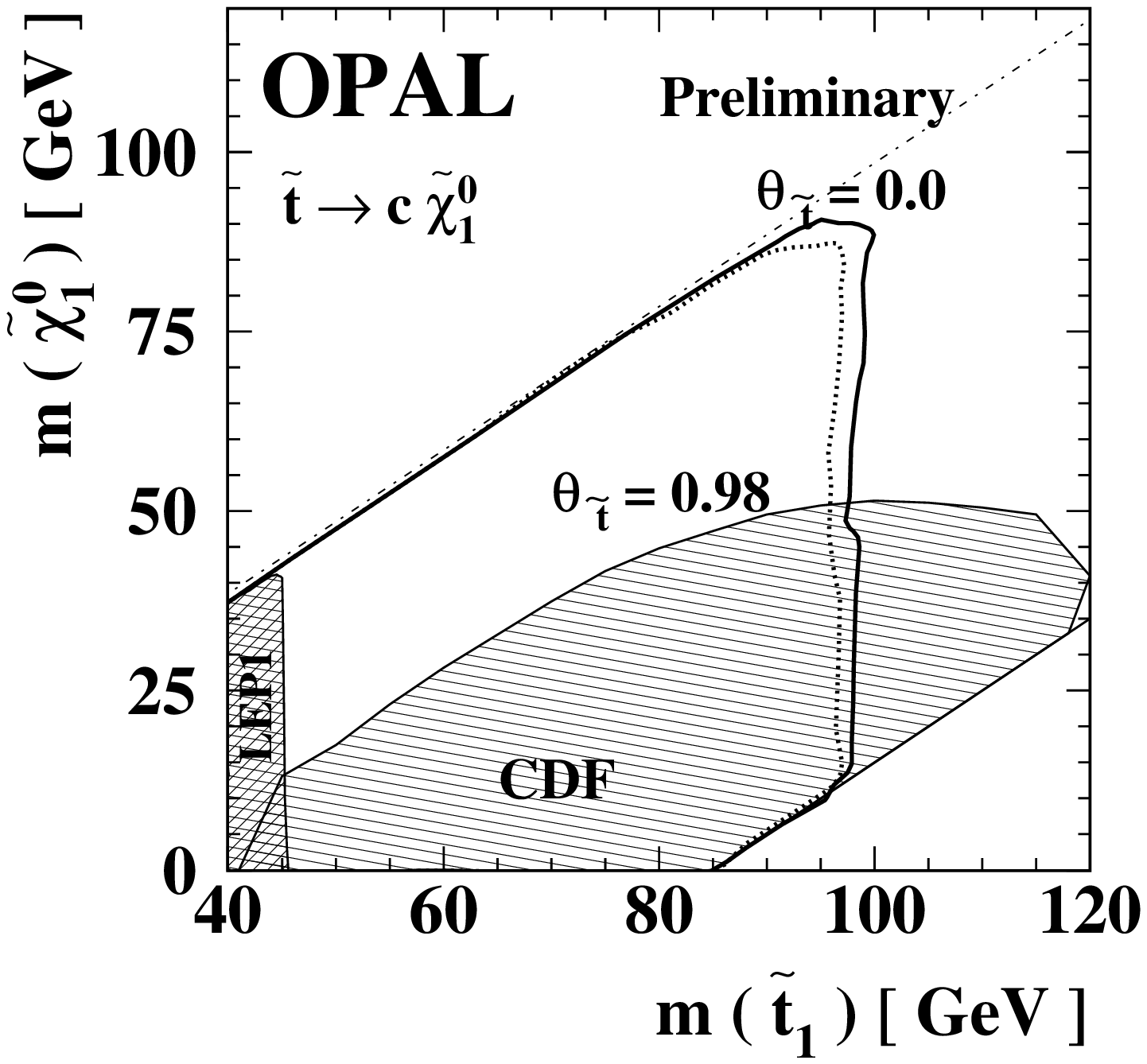,height=8.8cm}
\end{minipage}
\begin{minipage}{9.5cm}
\hspace*{-1.0cm}
\epsfig{file=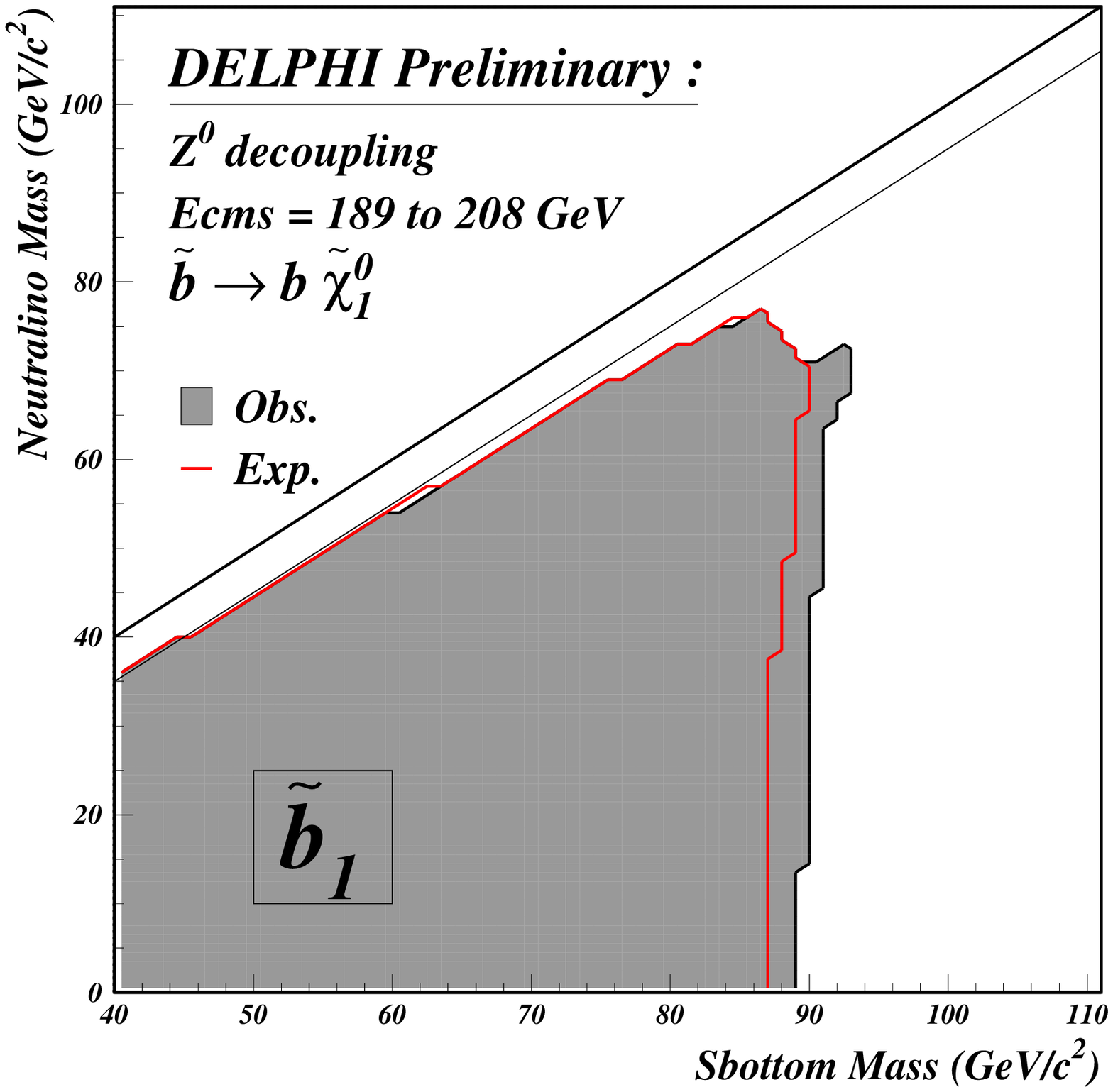,height=6.8cm}
\end{minipage}
\caption{Excluded regions (95\% C.L) in the 
$M_{\sq} - M_{\chione}$-plane for stops (left) and sbottoms (right). In the case of the 
stop the limit is shown for the no-mixing case (full line) and the case of 
minimal coupling
to the $Z^0$ (dotted line). 
For the sbottom exclusion 
also the expected limit is shown. 
\label{f:squarks}}
\end{figure}

Sbottom squarks will mainly decay into $b \chione$, generating events with two 
acoplanar b-jets and missing energy. At LEP the equivalent channel is not open 
for stops. Stop squarks can decay via loops into $c \chione$ or via a virtual 
chargino into $b l \sneu$, if the sneutrino is light enough. 
In Fig.\ref{f:squarks} examples of the exluded regions in the 
$( M_{\sq}-M_{\chione})$-plane are shown for 
stop (from the OPAL collaboration) and   
sbottom decays (from the DELPHI collaboration).
\begin{table}[h]
\caption{Limits (95\% C.L.) on the masses of sfermions from a combination of 
the data of the four LEP experiments.
\label{t:sfermions}}
\vspace{0.4cm}
\begin{center}
\begin{tabular}{| l | l  | c | l |}
\hline
      & Decay   &   Mass limit  &  condition    \\
      &     &   95\% C.L.   &               \\
\hline
  
& & & \\
Selectron  & $\sel_R \rightarrow e \chione$ & 99.4 GeV & $m_{\chione}$ = 40 GeV \\
Smuon      & $\smu_R \rightarrow \mu \chione$& 96.4 GeV& $m_{\chione}$ = 40 GeV \\
Stau       & $\stau_1 \rightarrow \tau \chione$ & 87.1 GeV  & $m_{\chione}$ = 40 GeV \\
\hline
        &         &      & \\
Stop    & $\tilde{t} \rightarrow c \chione$ & 95 GeV & $\delm$ = 40 GeV \\
        & $\tilde{t} \rightarrow c \chione$ &  65 GeV & independent  \\
        &                                   &         & of $\delm$  \\
        & $\tilde{t} \rightarrow b l \sneu$ & 97 GeV & $\delm$ = 40 GeV \\
Sbottom & $\sbot \rightarrow b \chione$ & 95 GeV & $\delm$ = 20 GeV \\
\hline
\end{tabular}
\end{center}
\end{table}
The sbottom exclusion is show for the 
decoupling case, the stop exclusion is given for both the no-mixing case and the 
case of minimal coupling to the $Z^0$. For low $\delm = M_{\tilde{t}} - M_{\chione}$ values
the stop lifetime becomes sizeable. The ALEPH collaboration has performed an analysis 
looking for stable stop particles and stop particles decaying with a sizeable 
lifetime~\cite{aleph}. 
This analysis allows to derive an absolute lower limit of 65 GeV on the mass of 
the stop, independent of $\delm$. The minimum is taken for a $\tanb$-values of 2.7. 
For larger $\delm$ values the limits on the stop and sbottom are significantly 
higher. The results obtained from the combination of the four LEP experiments 
are included in Table \ref{t:sfermions}. Depending on the decay mode, stop and sbottom
squarks are expected to be heavier than 95 to 97 GeV for a $\delm$ value of 40 GeV.  

\section{Searches for charginos and neutralinos}
\begin{figure}
\begin{minipage}{9.5cm}
\epsfig{file=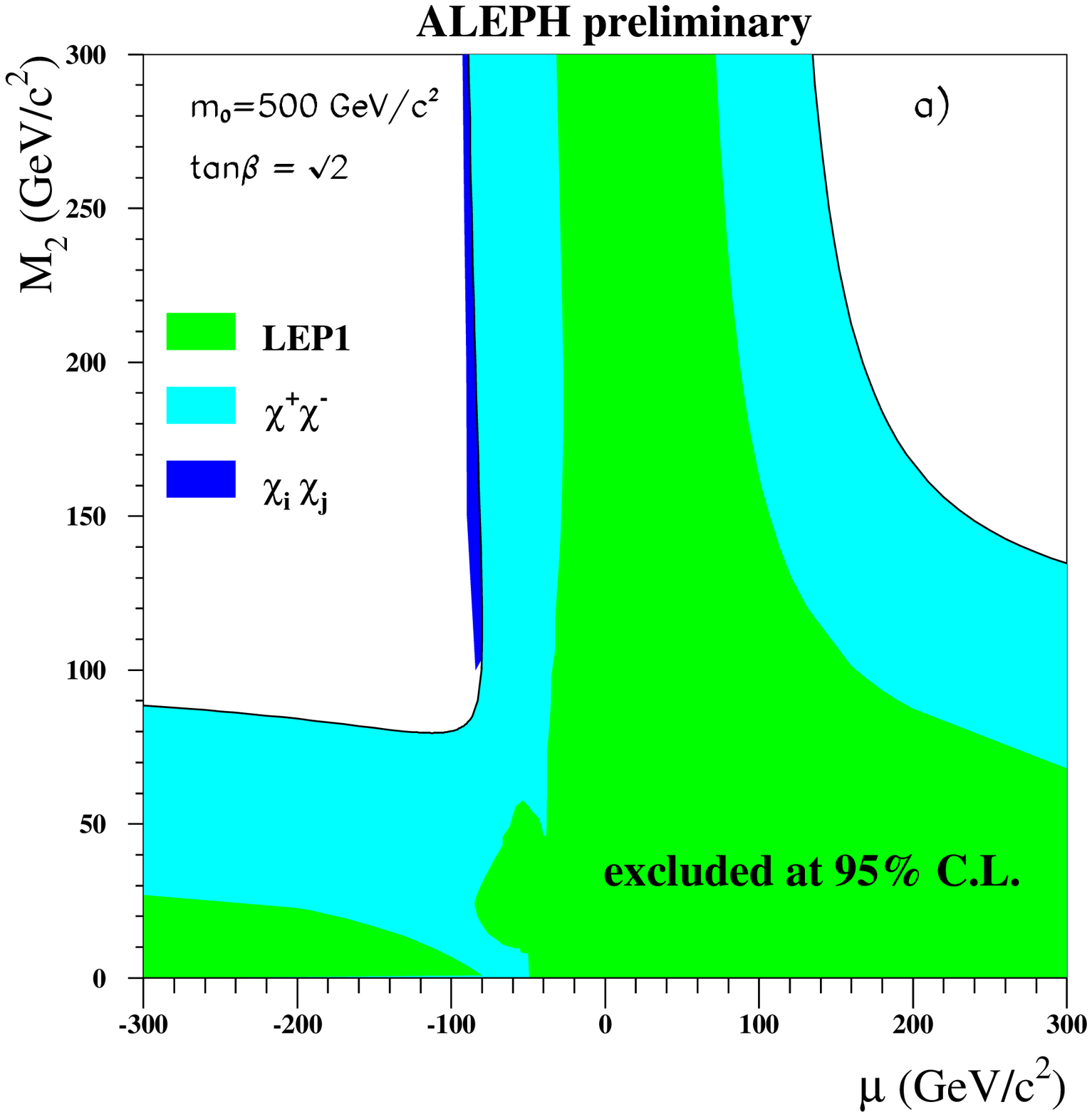,height=8.4cm}
\end{minipage}
\begin{minipage}{9.5cm}
\hspace*{-2.0cm}
\vspace*{0.4cm}
\epsfig{file=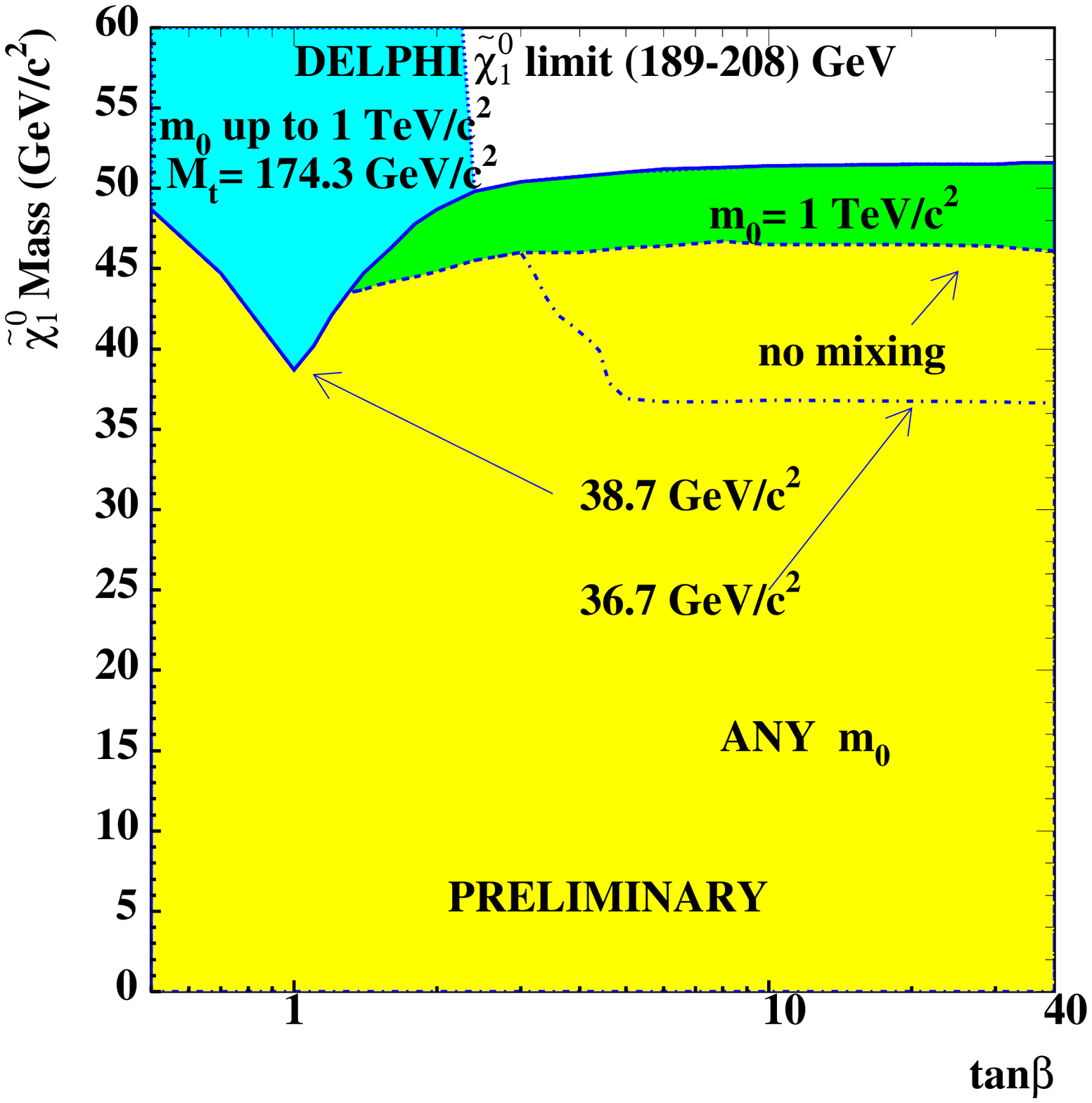,height=8.0cm}
\end{minipage}
\caption{(a) Excluded regions (95\% C.L) in the 
$\mumtwo$-plane for $m_0$ = 500 GeV and $\tanb$ = $\sqrt{2}$. The chargino searches 
cover regions nearly up to the kinematic limit ($\sim$ 103 GeV). 
Neutralino searches can exclude additional regions of parameter space 
in the higgsino region. 
(b) The lower limit (95\% C.L.) on the mass of the lightest neutralino as a function of 
$\tanb$. The solid curve shows the limit obtained for $m_0$ = 1000 GeV, the dashed curve
shows the limit obtained allowing for any $m_0$, assuming that there is no mixing in the 
third family. The dash-dotted curve shows the limit obtained for any $m_0$ assuming 
teh special mixing scenario with $A_{\tau} = A_{b} = A_{t} = 0$. The steep dashed curve 
shows the effect of the searches for the Higgs boson for the maximal $m_h$ scenario 
(no mixing scenario), $m_0 < 1000$ GeV and $m_{top}$ = 174.3 GeV. 
\label{f:chargino}}
\end{figure}

At LEP charginos can be pair produced via a $Z^0$ or a photon exchange in the 
s-channel or a sneutrino exchange in the t-channel. For large sfermion masses 
they predominantly decay via a $W^*$ into a $ff' \chione$ final state, which 
result in the detector in four jets, two jets and a charged lepton or two charged 
leptons, and missing energy due to neutralinos and neutrinos. Due to the 
t-channel contribution chargino mass limits depend on the sneutrino and the 
field content of the lightest chargino. Mass limits on the charginos are usually 
used to exclude regions in the  $\mumtwo$-plane, as shown in Fig.\ref{f:chargino}
for the case where $\tanb$ = 2.0 and a universal scalar mass parameter $m_0$ at the GUT 
scale of 500 GeV has been chosen.
The excluded parameter regions can still be enlarged beyond the kinematic limits
for chargino production by a search for pair production of neutralinos. 
Neutralino pair production proceeds via a Z-boson exchange in the s-channel and
via selectron exchange in the t-channel. Since pair production of the lightest 
neutralino is not directly observable one searches for events in which at least 
one of the heavier neutralinos is produced in association with the $\chione$. 
For smaller
values of the parameter $m_0$ which corresponds to a scenario with lighter 
sfermions and sneutrinos the production cross section for charginos is reduced, 
whereas the neutralino production cross section is increased. Given the light 
sfermion and sneutrino masses the leptonic branching ratio via light sfermions
also increases. Combining the chargino search with the search for sleptons
allows also in this case to exclude charginos with masses close to the 
kinematic limit. An analysis has been performed by the L3-collaboration~\cite{l3}, 
where the 
chargino limit is determined as a function of the sneutrino mass for a parameter point
at low $\tanb$ in the gaugino region ($\tanb$ = 2.0, $\mu$=-200 GeV). At this 
parameter point, charginos with masses below 98.6 GeV are excluded independent
of the sneutrino mass and thereby the parameter $m_0$. 

For very low mass differences $\delm$ below a few GeV,
the standard chargino analyses become inefficient. To cover also this case, 
dedicated analyses have been developed, which either search for events with an ISR 
photon or for events with heavy charged particles
passing through the detector. The latter search covers a possible long lifetime
scenario of the chargino, which appears if the chargino becomes nearly 
degenerate in mass with the neutralino. Such analyes have been performed by all
LEP collaborations. For large $m_0$ values, charginos with masses below 
about 90 GeV can be excluded, independent on the mass difference $\delm$ to 
the lightest neutralino. 

The excluded regions in the $\mumtwo$-plane can be compared with the $\chione$
iso-mass curves and can be translated into a lower limit on its mass. All LEP 
experiments have extracted limits for the large $m_0$ case. The lower 
mass limits extracted for the lightest neutralino  
are in the range between 38.0 GeV and 39.6 GeV. The LSP limit is 
essentially determined by the chargino searches, except for very low $\tanb$ values, 
where the neutralino searches improve the mass limit. The minimum as a function of 
$\tanb$ is found for $\tanb$ = 1.0.

The OPAL collaboration has presented a limit for the lightest neutralino independent 
of the $m_0$ value~\cite{OPAL}. The results of the chargino, neutralino and 
slepton searches have been combined and the SUSY parameters $M_2$, $\tanb$, $\mu$ and 
$A_0$ have been scanned in the range 
$ 0 < M_2 < 2000$  GeV, $|\mu| < 500$ GeV 
and $A_0 = 0, \pm M_2$. In this analysis the limit of the lightest neutralino is 
determined to be 36.3 GeV. 

A discussion of the LSP limit for any $m_0$ has also been presented by the 
DELPHI collaboration~\cite{DELPHI}. In Fig.~\ref{f:chargino} the lower limit for 
$m_{\chione}$ is shown as a function of $\tanb$ for any value of $m_0$. The 
$m_0$-independent limit, resulting from a combination of chargino, neutralino and 
slepton searches, follows the large $m_0$ limit up to $\tanb$= 1.4. Then a lower 
value is taken due to 
the opening of the chargino-sneutrino mass degeneracy. For large $\tanb$ values 
($\tanb > 5.0$) the LSP limit takes its lowest value of 36.7 GeV 
due to the mass degeneracy between
the lightest stau and the lightest neutralino. In this case mixing in the stau sector is taken 
into account by setting the parameter $A_{\tau}$, which appears in the off-diagonal 
matrix element of the stau mass matrix ($A_{\tau} - \mu \tanb$), to zero. 
For a more detailed discussion of the stau mixing effects the reader is referred to 
Ref.~\cite{DELPHI}.

\section{Conclusions}

LEP2 has made significant contributions in the exploration of the SUSY landscape. Up to 
the highest LEP2 energies around 208 GeV no evidence for the production of SUSY 
particles has been found. Stringent limits on the masses of sfermions, charginos and 
the lightest neutralino have been set in the framework of the MSSM, assuming 
R-parity conservation and the lightest neutralino to be the lightest supersymmetric 
particle. The analyses of the final LEP data are already well advanced.  Future 
analyses will concentrate on more complete interpretations, including, for example, 
stau mixing effects, and on the combination of the data of all LEP experiments.

\section*{Acknowledgments}
It is a pleasure to thank the many colleagues working in the SUSY 
working groups of the LEP collaborations for their hard work to extract these 
results on the short timescale since the termination of the LEP runnning.

\end{document}

%% file: kjtex.tex
\newcommand{\pbs}{\mbox{$\rm{pb}^{-1}$}}

\newcommand{\chione}{\mbox{$\tilde{\chi}^0_1$}}

\newcommand{\sel}{\mbox{$\tilde{e}$}}
\newcommand{\smu}{\mbox{$\tilde{\mu}$}}
\newcommand{\stau}{\mbox{$\tilde{\tau}$}}
\newcommand{\slep}{\mbox{$\tilde{l}$}}

\newcommand{\sneu}{\mbox{$\tilde{\nu}$}}

\newcommand{\sq}{\mbox{$\tilde{q}$}}

\newcommand{\sbot}{\mbox{$\tilde{b}$}}

\newcommand{\tanb}{\mbox{$\tan \beta$}}

\newcommand{\delm}{\ensuremath{\Delta M}}

\newcommand{\mumtwo}{\ensuremath{(\mu - M_2)}}